\documentclass[nofootinbib,twocolumn,showpacs,preprintnumbers,pre,aps]{revtex4-1}

\usepackage{graphicx}
\usepackage{bm}
\usepackage{amsmath}
\usepackage{amssymb}
\usepackage{color}

\begin{document}

\title{Budding transition of asymmetric two-component lipid domains}%

\author{Jean Wolff}\email{jean.wolff@ics-cnrs.unistra.fr}

\affiliation{Institut Charles Sadron, UPR22-CNRS 23,
rue du Loess BP 84047, 67034 Strasbourg Cedex, France}

\author{Shigeyuki Komura}\email{komura@tmu.ac.jp}

\affiliation{Department of Chemistry, Graduate School of Science and Engineering,
Tokyo Metropolitan University, Tokyo 192-0397, Japan}

\author{David Andelman}\email{andelman@post.tau.ac.il}

\affiliation{Raymond and Beverly Sackler School of Physics and Astronomy,
Tel Aviv University, Ramat Aviv, Tel Aviv 69978, Israel}
\affiliation{CAS Key Laboratory of Theoretical Physics, Institute of Theoretical Physics, Chinese
Academy of Sciences,
Beijing 100190, China}

\date{August 19, 2016}

\begin{abstract}
We propose a model that accounts for the budding transition of asymmetric two-component lipid domains, where the two monolayers (leaflets) have different average compositions controlled by
independent chemical potentials.
Assuming a coupling between the local curvature and  local lipid composition in each of the leaflets, we discuss the morphology and thermodynamic behavior of asymmetric lipid domains.
The membrane free-energy contains three contributions: the bending energy, the line tension, and a Landau free-energy for a lateral phase separation.
Within a mean-field treatment, we obtain various phase diagrams containing fully budded, dimpled, and flat states as a function of the two leaflet compositions.
The global phase behavior is analyzed, and depending on system parameters, the phase diagrams include one-phase, two-phase and three-phase regions.
In particular, we predict various phase coexistence regions between different morphologies of domains, which may be observed in multi-component membranes or vesicles.
\end{abstract}

\maketitle

\section{Introduction}
\label{sec:introduction}

The cytoplasmic membrane separates the living cell from its extra-cellular surroundings,
while other intra-cellular membranes compartmentalize cellular organelles.
Biomembranes are constructed from two monolayers (leaflets) in a back-to-back arrangement, and are in general asymmetric in their lipids composition~\cite{Devaux,Meer}. For example, in human red blood cells, the inner cytoplasmic leaflet is composed mostly of phosphatidylethanolamine (PE) and phosphatidylserine (PS), while the outer cytoplasmic
leaflet is composed of phosphatidylcholine (PC), sphingomyelin (SM) and a variety of glycolipids~\cite{Opden,Devaux1}.
The asymmetric nature of the cell membrane plays a key role in a variety of cellular processes such as endocytosis~\cite{Pomo}, vesicle budding and trafficking~\cite{Dal1}. Furthermore, in living cells, the composition asymmetry is an active and energy-consuming process. It is maintained by several membrane proteins such as flippase and floppase that allow lipids to exchange between the two leaflets with the aid of adenosine triphosphate (ATP)~\cite{Dal}.

In artificial multi-component lipid bilayers, the two
membrane leaflets can undergo a lateral phase separation. Several authors made
the connection between such a phase separation in artificial membranes
and existence of small dynamic domains (``rafts") in biological
membranes~\cite{KA14}. It should be noted, however, the size of rafts
in biological membranes are expected to be in the range of 10--100~nm~\cite{Simons,Munro}.
Raft are believed to be enriched mixtures of cholesterol and SM in a liquid-ordered phase
(L$_{\rm o}$), embedded in a background of a liquid-disordered phase (L$_{\rm d}$). Despite the lack of an ultimate proof for the existence of rafts, they have been advocated in relation with their potential influence on biological cellular processes.
It has been suggested that rafts act as organizing centers for the
assembly of signaling molecules, influencing membrane fluidity, and
regulating receptor trafficking~\cite{Simons,Munro}.

\begin{figure*}[tbh]
\begin{center}
\includegraphics[scale=0.4]{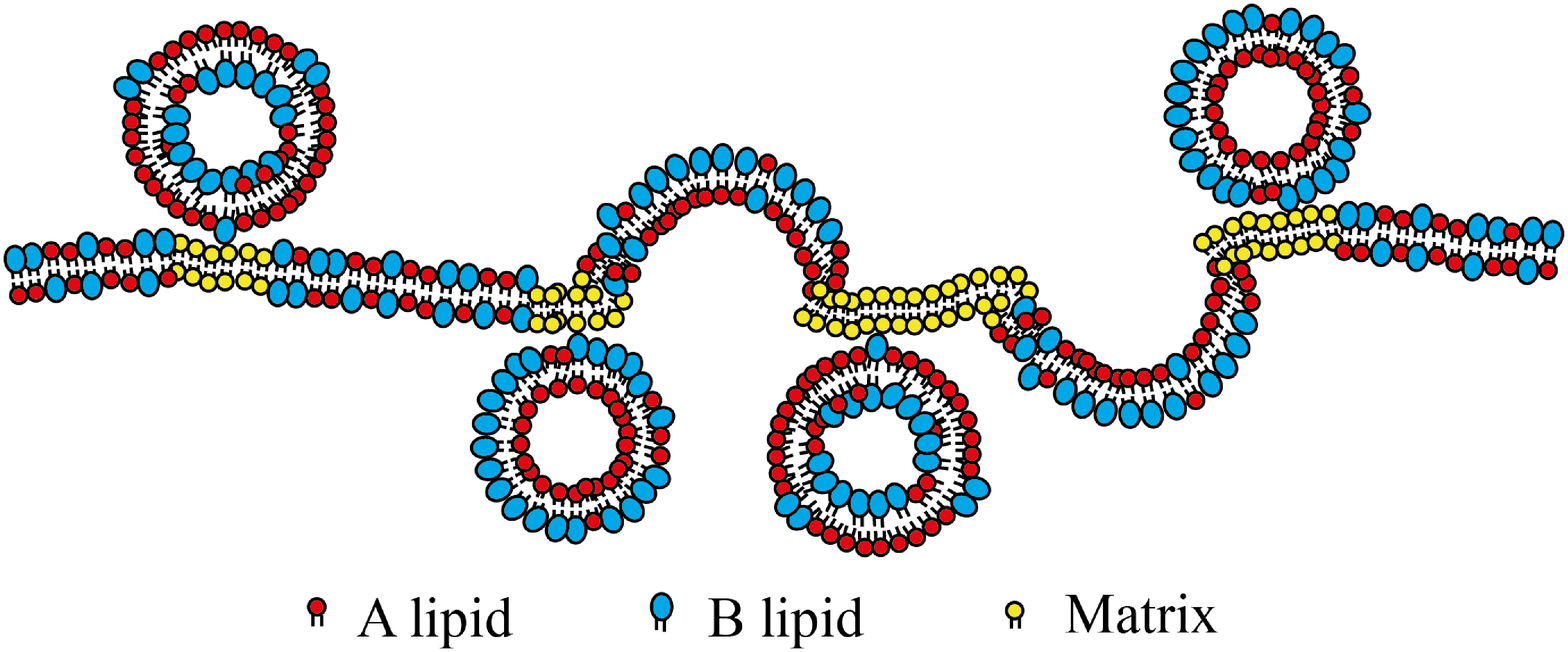}
\end{center}
\caption{\textsf{
Schematic vertical cut through a membrane consisting of several domains.
The domains are embedded in a lipid matrix (yellow) and have a line tension acting
along the domain rim.
Three domain types can be seen and are further explained in Fig.~2: Flat (F), Budded (B)
and Dimpled (D).
Each domain is formed by two lipids (red and blue) that  partition themselves differently
on the two domain leaflets.
}}
\label{fig1}
\end{figure*}

A theoretical model for domain-induced budding of planar membranes was proposed by
Lipowsky~\cite{Lipowsky92} some years ago, and later was extended for closed
vesicles~\cite{JL93,JL96}.
In the model, the competition between the membrane bending energy and domain line tension leads to a budding transition.
More recently, a model describing domain-induced budding in bilayers composed of a binary mixture of lipids was proposed by us~\cite{Wolff}.
In particular, we have shown that dimpled domains are formed and remain stable due to an
asymmetry between the two compositions of the corresponding domain leaflets,
given that the line tension along the domain rim is not too large.
The calculations in Ref.~\cite{Wolff} were done for a specific case where
the relative concentration between the two lipids on the bilayer domain stays constant, while
the lipids are allowed to freely exchange between the two leaflets.

In the last decade, however, techniques such as Langmuir-Blodgett or
Langmuir-Sch\"{a}efer have enabled a control over the asymmetric composition
of artificial membranes~\cite{Kiessling, Cheng}.
For example, unsupported bilayers via the Montal-M\"uller method have been used to form asymmetric bilayers in which the composition of each leaflet was independently controlled~\cite{Collins}. A key ingredient in understanding those experiments is the fact that
the flip-flop process that exchanges lipids across the leaflets is slower than experimental times.
Given these experimental findings, it is worthwhile to consider bilayers where each of the leaflet domain composition (rather than the overall bilayer composition) can be controlled in a separate and independent way.

In this paper, we generalize our previous budding model~\cite{Wolff} and extend it to asymmetric two-component lipid domains.
Each domain leaflet has a conserved lipid composition that is independent from that of the other leaflet.
We consider the possibility of domains curved in the third dimension, which can produce buds as shown in Fig.~\ref{fig1}.
In our model, the composition-dependent spontaneous curvature leads to coupling between curvature and lipid composition in each of the domain leaflets~\cite{Leibler,SPA,MS}.
Such a coupling leads to rich phase behavior including various phase coexistence regions.

For the sake of clarity, we do not take into account any direct interaction between two domains that face each other, although various possibilities have been previously proposed~\cite{May}. In addition, we would like to mention that in a separate set of studies for completely planar membranes, asymmetric bilayers composed of two modulated monolayers
(leaflets) were considered~\cite{HKA09,HKA12}. In these works, the static and dynamic properties of concentration fluctuations have been presented together with the related micro-phase separation.

Our model may have several experimental implications.  We predict that bilayer domains can exist in three states having different equilibrium shapes: fully
budded, dimpled, and flat states. Their relative stability depends on controlled system parameters: temperature, degree of compositional asymmetry
between the two leaflets and domain size.
We find that the dimpled state is the most stable one in some of the parameter range and this is in accord
with recent experiments~\cite{RUPK}.
Based on the calculated phase diagrams, we anticipate that membranes should exhibit in some parameter range two-phase
and three-phase coexistence between different domain states.

The outline of this paper is as follows.
Section~\ref{sec:model} generalizes our previous budding model~\cite{Wolff}, and the free-energy describing asymmetric two-component lipid domains is discussed. In Sec.~\ref{sec:diagram}, we explain the conditions for various phase equilibria and how to calculate the phase diagrams.
We then proceed by presenting the phase diagrams in Sec.~\ref{sec:result}, and discuss the resulting global phase behavior.
Finally, a more qualitative discussion is provided in Sec.~\ref{sec:discussion}.

\section{Model}
\label{sec:model}

\begin{figure}[tbh]
\begin{center}
\includegraphics[scale=0.4]{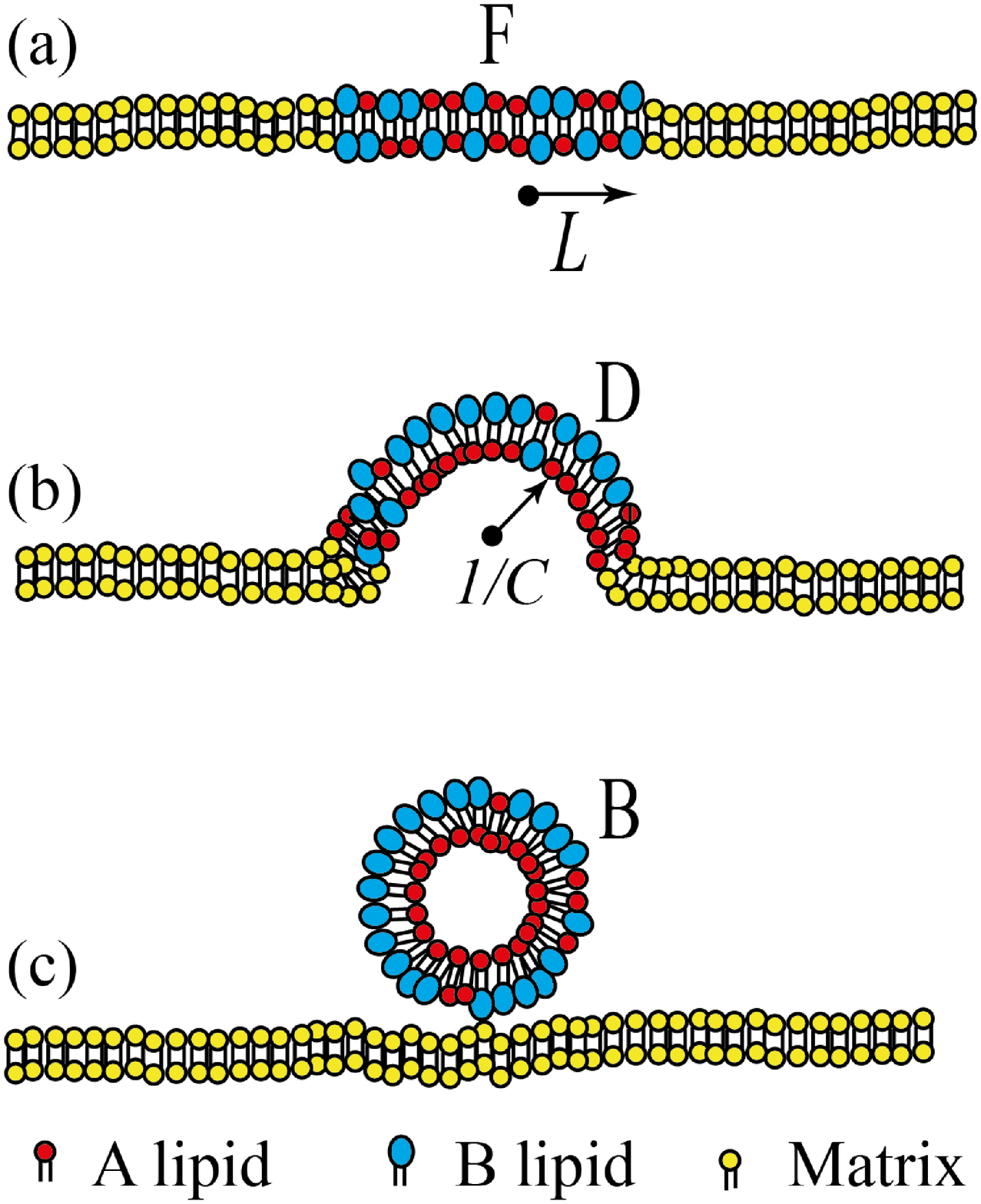}
\end{center}
\caption{\textsf{
A schematic vertical cut through three possible domain shapes:
(a) a flat bilayer domain composed of a mixture of A and B lipids (red and blue, respectively), and embedded in an otherwise flat membrane (yellow).
The circular flat domain (F) has a radius $L$ and area $S=\pi L^2$.
(b) A partial bud (dimpled domain) curved in the out-of-plane direction.
The bud (D) of the same area $S$ forms a spherical cap of radius $1/C$, where $C$ is the curvature. (c) A fully-budded domain (B) has a spherical shape of total area $S$, and just touches the flat membrane. The line tension $\gamma$ acts along the line boundary between the domain (red/blue lipids)
and the flat membrane matrix (yellow lipids).
}}
\label{fig2}
\end{figure}

We consider a membrane consisting of two monolayers (the terms ``monolayer" and ``leaflet" will be used interchangeably in this paper), each composed of an A/B mixture of lipids, which partition themselves asymmetrically between the two leaflets.
We assume that the membrane can undergo a lateral phase separation creating domains of different lipid composition. The domains are taken to fully span the two monolayers, but
the leaflet compositions in these domains can be different.  Hence, the formed lipid domains are, in general, asymmetric.
We further assume that the two leaflet compositions are conserved and can be taken as independent from each other.
In the following, we discuss the thermodynamic behavior of a system consisting of a large number of such asymmetric domains as shown in Fig.~\ref{fig1}. For this purpose, we explain below the different terms that contribute to the domain free-energy.

In Fig.~\ref{fig2}, we show a vertical cut through three possible domain states: flat (F), dimpled (D) and fully-budded (B).
The flat circular domain (F) in (a) has an area,  $S=\pi L^2$, which is assumed to remain constant during the budding process.
For simplicity, we consider in (b) dimpled buds (D) whose shape is a spherical cap of radius $1/C$,  and in (c) the extreme case of a completely detached spherical bud (B).
The total bending energy of the budded domain is given by the curvature
contributions from its two monolayers~\cite{Helfrich73,SafranBook}:
\begin{equation}
E_{\rm bend} =2 \pi L^2 \kappa \left[ (C -C_0)^2
+ (C + C_0)^2 \right],
\label{bend}
\end{equation}
where $\kappa$ is the bending rigidity modulus (assumed to be a constant that is independent of lipid composition) and $C_0$ the monolayer spontaneous curvature. As shown in Fig.~\ref{fig2}, the two monolayer deform with curvatures $+C$ and $-C$, respectively.

The second energy contribution is the domain edge energy that is proportional to the perimeter length and to the line tension, $\gamma$~\cite{Lipowsky92}:
\begin{equation}
E_{\rm edge} = 2 \pi L \gamma\sqrt{1-(LC/2)^2}.
\label{edge}
\end{equation}
Note that in the extreme case, when the domain buds into a complete spherical shape (B) as in Fig.~\ref{fig2}(c), $C=\pm 2/L$ and $E_{\rm edge}=0$.
In the above, the strong variation in composition between the domain and its surrounding matrix is effectively taken into account through the line tension, $\gamma$, which is treated as an external control parameter.
This situation can be justified for a strong segregation that results
in a sharp boundary between the domain and its flat matrix surroundings.

As each domain is composed of an A/B mixture, we define
$\phi_{\rm A}$ ($\phi_{\rm B}$) as the area fraction (assumed to be equal to the molar fraction) of the
${\rm A}$ lipid ($\rm B$ lipid) in the upper leaflet domain, and similarly,
$\psi_{\rm A}$ ($\psi_{\rm B})$ for the lower leaflet domain.
We assume that each monolayer is incompressible so that
$\phi_{\rm A}+\phi_{\rm B}=1$ and $\psi_{\rm A}+\psi_{\rm B}=1$.
Hence, the two relevant order parameters are the  relative composition
in the upper leaflet:
\begin{equation}
\phi= \phi_{\rm A}-\phi_{\rm B} ,
\end{equation}
and in the lower leaflet,
\begin{equation}
\psi= \psi_{\rm A}-\psi_{\rm B} .
\label{def}
\end{equation}
As in any A/B mixture, the possibility of a phase separation can be described by a phenomenological Landau expansion of the free-energy in powers of $\phi$ and $\psi$. This expansion is done separately for
each monolayer, and the total contribution
to the free-energy is the sum over the two monolayers:
\begin{equation}
E_{\rm phase} = \pi L^2 \frac{U}{\Xi^2}
\left[ \frac{t}{2} (\phi^2 +\psi^2)+ \frac{1}{4} (\phi^4+\psi^4) \right],
\label{phase}
\end{equation}
where $\Xi \equiv \kappa/\gamma$ is the invagination length,
$U$ is a parameter that sets the energy scale, and
$t \sim (T-T_{\rm c})/T_{\rm c}$  is the reduced temperature ($T_{\rm c}$ being the critical temperature).
In Eq.~(\ref{phase}) above, we multiply by the domain area, $\pi L^2$, to obtain the domain free-energy.

Hereafter, we will use several dimensionless variables: a rescaled curvature $c \equiv LC$, rescaled spontaneous curvature $c_0 \equiv LC_0$, and rescaled invagination length $\xi\equiv \Xi/L$.
The coupling between curvature and composition is taken into account by
assuming a linear dependence of the spontaneous curvature $c_0$ on the relative composition in each of the leaflets~\cite{Leibler,SPA,MS}:
\begin{equation}
c_0(\phi) =\bar{c}_0-\beta \phi,
\label{spontaneous1}
\end{equation}
\begin{equation}
c_0(\psi) =\bar{c}_0-\beta \psi,
\label{spontaneous2}
\end{equation}
where $\bar{c}_0$ is the monolayer spontaneous curvature for the symmetric 1:1 composition,
$\phi=\psi=0$, and $\beta$ is a coupling parameter that has the same value for the two monolayers.
Since $\bar{c}_0$ is a constant, it merely shifts the origin of the chemical potential, and will be dropped out without loss of generality.

The total free-energy per domain is then given by the sum of Eqs.~(\ref{bend}), (\ref{edge}) and (\ref{phase}): $E_{\rm tot}=E_{\rm bend}+E_{\rm edge} + E_{\rm phase}$, and
its dimensionless form, $\varepsilon=E_{\rm tot}/2\pi \kappa$, is expressed as
\begin{align}
\varepsilon(\phi, \psi, c)& =(c+\beta\phi)^2+(c-\beta\psi)^2
+\frac{1}{\xi}\sqrt{1-c^2/4} \nonumber \\
&+\frac{1}{\xi^2} \left( \frac{U}{2 \kappa} \right)
\left[\frac{t}{2}(\phi^2+\psi^2)+\frac{1}{4}(\phi^4+\psi^4)
\right].
\label{eq6}
\end{align}
We note that Eq.~(\ref{eq6}) depends on three dimensionless parameters: $\beta$,
$\xi$, and $U/(2 \kappa)$, while the thermodynamic variables are the reduced temperature
$t$ and the three order parameters: $\phi$, $\psi$ and $c$.
In the calculations presented hereafter, we set $U/(2 \kappa)=1$ and vary the
values of $\beta$ and $\xi$.

Within mean-field theory, the equilibrium states and phase transitions
are determined by minimizing $\varepsilon$ with respect to $\phi$, $\psi$ and $c$, under the condition that $\phi$ and $\psi$ are conserved order parameters while $c$ is not. From the minimization of $\varepsilon$ with respect to $c$, we obtain the condition
\begin{equation}
2(c+\beta\phi)+2(c-\beta\psi)-\frac{c}{4\xi\sqrt{1-c^2/4}}=0.
\label{eq7}
\end{equation}
Then, by substituting the curvature $c=c(\phi, \psi)$ into Eq.~(\ref{eq6}), results in a partially minimized free-energy, $\varepsilon^*$
\begin{equation}
\varepsilon^{\ast}(\phi, \psi) = \varepsilon(\phi, \psi, c(\phi, \psi)),
\label{epsast}
\end{equation}
as a function of $\phi$ and $\psi$.

Typical experimental values of domain sizes are in the range of $L\simeq 50$--$500$~nm~\cite{Simons}, the bending rigidity
$\kappa \simeq 10^{-19} {\rm J}$~\cite{SafranBook},
and  the line tension
$\gamma \simeq 0.2$--$6.2 \times 10^{-12}$~J/m~\cite{baumgart,tian}.
Hence, the scaled invagination length is estimated to be in the range, $\xi \simeq 0.01$--$10$. These values will be used in the next section where we calculate numerically the phase diagrams.

\section{Phase equilibria conditions}
\label{sec:diagram}

In order to obtain various phase coexistence regions, $\varepsilon^{\ast}$ in Eq.~(\ref{epsast}) should be further minimized with respect to the conserved order parameters, $\phi$ and $\psi$. Hence, we consider the following thermodynamical potential
\begin{equation}
g(\phi, \psi)= \varepsilon^{\ast}(\phi,\psi)-\mu_{\phi}\phi-\mu_{\psi}\psi,
\label{grand}
\end{equation}
where $\mu_{\phi}$ and $\mu_{\psi}$ are the chemical potentials coupled with the A/B
relative compositions in the upper and lower domains, respectively.
They act as Lagrange multipliers that take into account the conserved $\phi$
and $\psi$ compositions.
In general, these two chemical potentials have different values, $\mu_{\phi} \ne \mu_{\psi}$.
The special case of $\mu_{\phi} = \mu_{\psi}$, for which only the total relative composition,
$\phi + \psi$, is conserved was investigated in our previous work~\cite{Wolff}, while here we
deal with a general situation where each of the compositions, $\phi$ and $\psi$,
are conserved independently.

\begin{figure}[tbh]
\begin{center}
\includegraphics[scale=0.4]{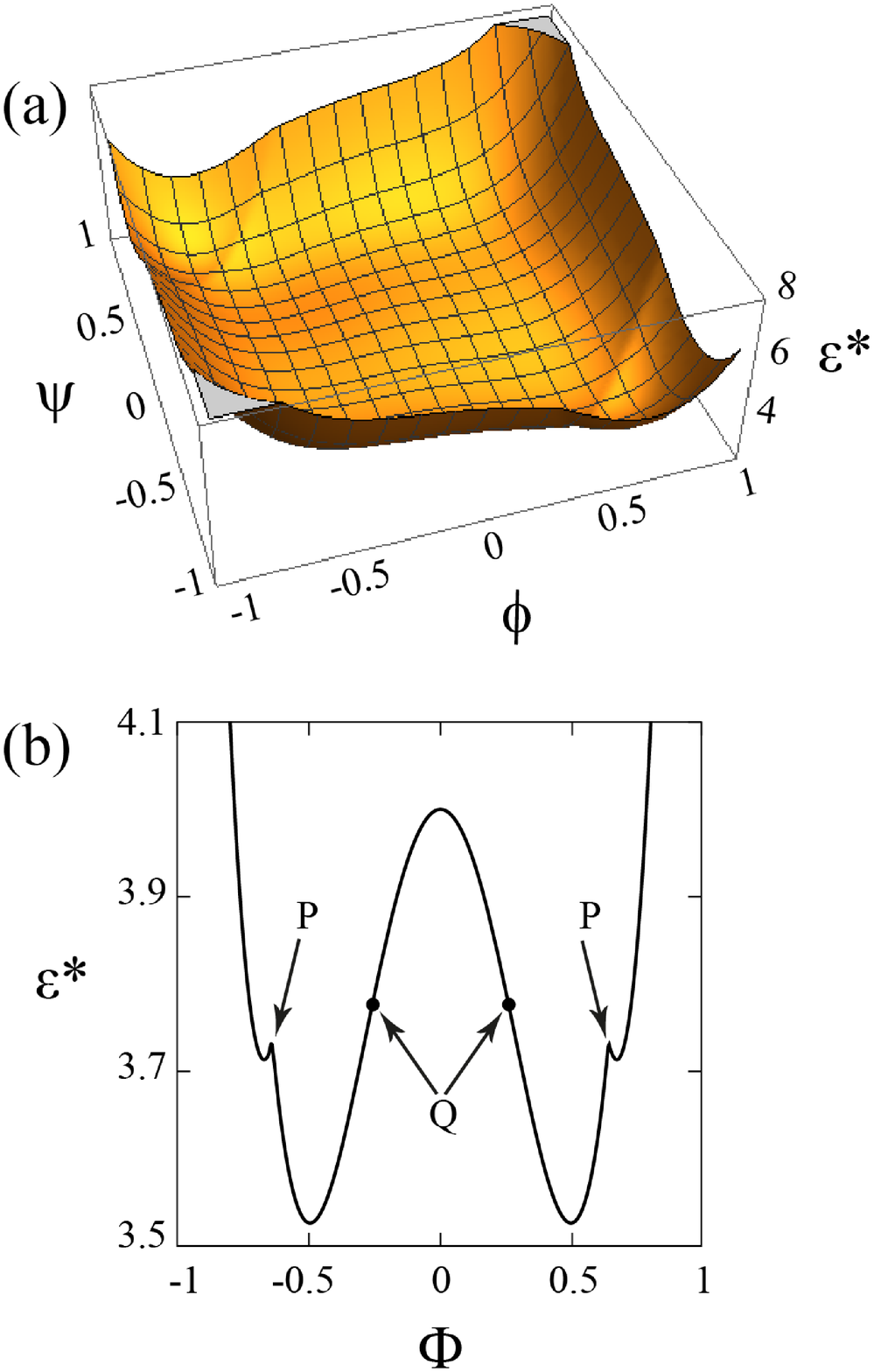}
\end{center}
\caption{\textsf{
(a) Plot of the partially minimized free-energy $\varepsilon^{\ast}(\phi, \psi) $ as a function of $\phi$ and
$\psi$ for $\xi=0.25$, $\beta=1$ and $t=-0.2$ (below $T_{\rm c}$).
(b) A cut through the free-energy landscape $\varepsilon^{\ast}$ in the particular direction,
$\phi+\psi=0$, plotted as a function of $\Phi = (\phi-\psi)/2$.
The two P points are cusps at which the slope of $\varepsilon^*$ changes discontinuously, although the Hessian $H$ remains positive (see Appendix).
The two Q points correspond to the location at which $H$ vanishes.
}}
\label{fig3}
\end{figure}

The thermodynamic equilibrium between the two coexisting phases denoted as `1' and `2' and characterized by $(\phi_1, \psi_1)$ and $(\phi_2, \psi_2)$, satisfies the conditions~\cite{Gibbs}
\begin{eqnarray}
\left. \partial_\phi g(\phi,\psi)\right|_1&=&
\left. \partial_\phi g(\phi,\psi)\right|_2=0,\nonumber\\
\left. \partial_\psi g(\phi, \psi)\right|_1&=&
\left. \partial_\psi g(\phi,\psi)\right|_2=0,\nonumber\\
g(\phi_1, \psi_1)&=& g(\phi_2, \psi_2).
\label{eq123}
\end{eqnarray}
Similarly, for a three-phase coexistence between phases `1', `2' and `3', the following set of conditions should be satisfied:
\begin{eqnarray}
\label{3phase}
\left. \partial_\phi g(\phi,\psi)\right|_1&=&
\left. \partial_\phi g(\phi, \psi)\right|_2=
\left. \partial_\phi g(\phi, \psi)\right|_3=0\nonumber\\
\left. \partial_\psi g(\phi,  \psi) \right|_1&=&
\left. \partial_\psi g(\phi, \psi) \right|_2=
\left. \partial_\psi g(\phi, \psi) \right|_3=0\nonumber\\
g(\phi_1, \psi_1)&=& g(\phi_2, \psi_2)=g(\phi_3, \psi_3).
\end{eqnarray}

In Fig.~\ref{fig3}(a), we show an example of the partially minimized free-energy, $\varepsilon^{\ast}(\phi,\psi)$, as a function of $\phi$ and $\psi$ at a fixed temperature $t=-0.2$ (below $T_{\rm c}$), and for given
values of $\xi$ and $\beta$. In order to have a better view of the free-energy surface, we show in Fig.~\ref{fig3}(b) a
cross-section cut of the free-energy surface in the direction of $\Phi = (\phi-\psi)/2$, while keeping $\phi+\psi=0$.
Here, we see two singular cusps (points P) where the determinant of the Hessian matrix, $H$ (see Appendix) does not vanish. At these cusps, $H$ changes discontinuously although it remains positive. At points Q, on the other hand, the Hessian $H$ vanishes.
More details on the Hessian matrix and determinant, and their relation to the phase stability and spinodal lines are presented in the Appendix.
By calculation the Hessian $H$, it is possible to derive the spinodal lines and critical points. Note that in most cases, the critical points and spinodal will not be shown on the phase diagrams, because they are preempted by the first-order phase transition lines and coexistence regions.

The phase diagrams are obtained by further minimizing $\varepsilon^{\ast}$, with respect to the two independent variables, $\phi$ and $\psi$. Convex regions of the free energy
correspond to single thermodynamical phases. Two-phase coexistence regions correspond to non-convex regions, where we can
construct a common tangent plane. The plane touches the free-energy surface at two points that determine the two phases in coexistence.
A more special three-phase coexistence region corresponds to a plane that touches the free-energy surface at three points.
More details on the numerical procedure of finding the phase diagrams are given below.

The numerical computation of the phase diagram is performed using a public-domain software
called ``Qhull"~\cite{qhull}.
The Qhull software generates initially a fine grid of triangulation,
which approximates the free-energy surface,
$\varepsilon^{\ast}$.
The three-phase coexistence regions correspond to facets with all sides being much larger than
the initial discretization.
The two-phase coexistence regions are associated with elongated triangles having one short side
that is much smaller than the other two longer sides that approximate the tie-lines.
Finally, small triangular facets of the free-energy surface are associated with stable one-phase regions.
The projection of the triangulated free-energy surface onto the composition plane provides
a systematic approximation for the phase diagram.
The Qhull results are then used as an initial condition in calculating more precisely
the equilibrium phase diagrams, including the various phase coexistence, Eqs.~(\ref{grand})--(\ref{3phase}).

\begin{figure}[tbh]
\begin{center}
\includegraphics[scale=0.5]{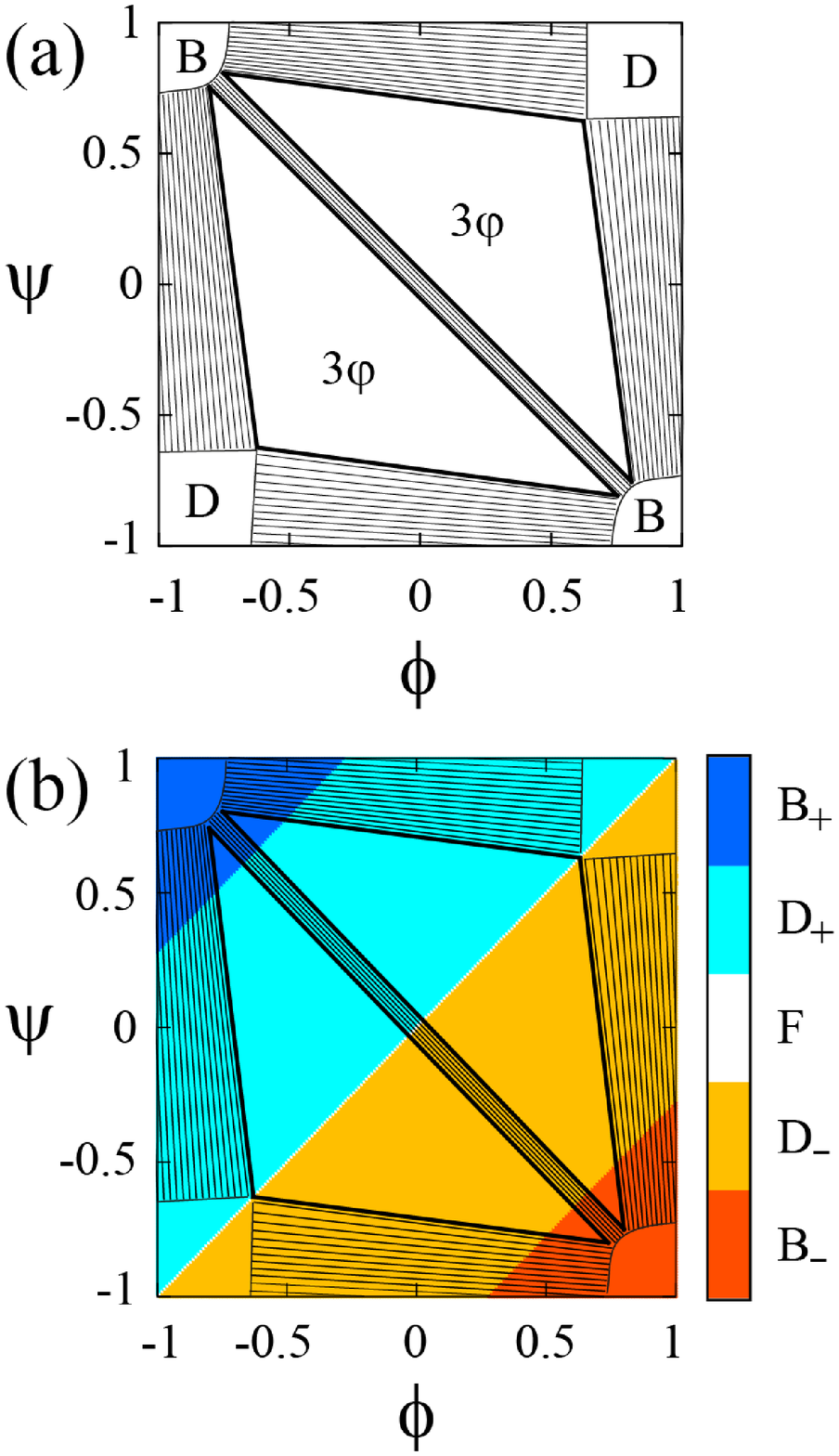}
\end{center}
\caption{\textsf{
(a) Phase diagram in the ($\phi$, $\psi$) plane for $\beta=1$, $\xi=0.25$ and $t=-0.45$.
The corners of the phase diagram indicates the four one-phases: D$_\pm$ and B$_\pm$, while
the flat F phase (not drawn) strictly lies only on the diagonal $\phi=\psi$.
The black lines represent tie-lines in the two-phase regions, and the two triangles are the
three-phase regions (see text for more details).
(b) The phase diagram is plotted as in (a) but with a superimposed colored plot for the
curvature, $c$.
As one crosses the major diagonal, $\phi=\psi$,
there is a smooth change from  D$_+$ ($c>0$) through the flat F ($c=0$)
to the D$_-$ ($c<0$).
Furthermore, the curvature also changes smoothly inside
the $D_\pm$ phases, but
the gradient in orange (D$_-$) and light blue (D$_+$) colors is not shown for clarity. However, the curvature
has a jump between B$_-$ ($c=-2$, red) and  D$_-$ ($-2< c <0$, orange) regions, as well as
between B$_+$ ($c=2$, dark blue) and  D$_+$ ($0< c<2$, light blue) ones.
}}
\label{fig4}
\end{figure}

\begin{figure}[tbh]
\begin{center}
\includegraphics[scale=0.5]{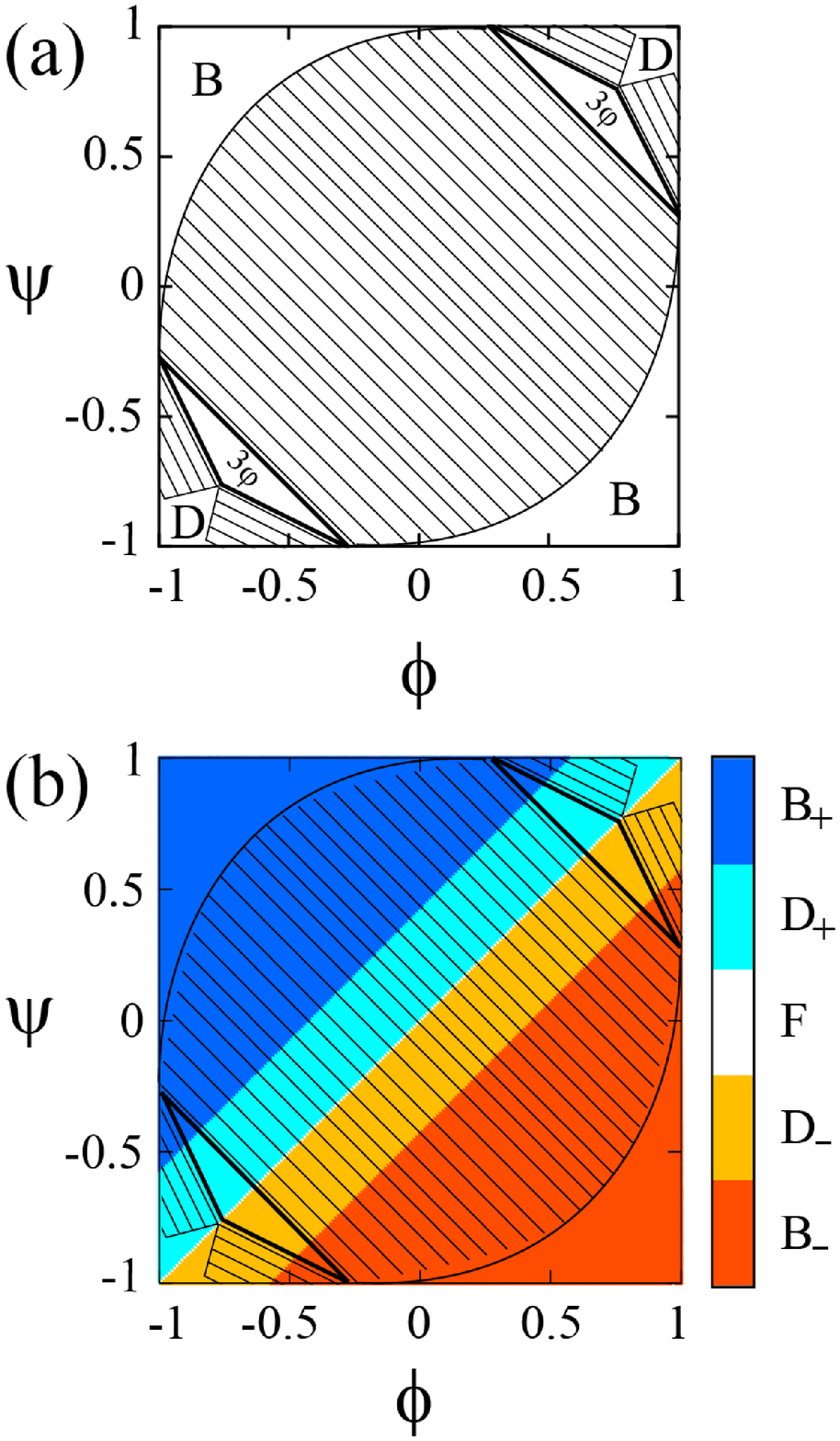}
\end{center}
\caption{\textsf{
(a) Phase diagram in the ($\phi$, $\psi$) plane as in Fig.~\ref{fig4} but with
 $\beta=3$, $\xi=0.25$ and $t=-0.45$.
(b) The phase diagram is plotted as in (a) but with a superimposed colored plot for the
curvature, $c$.
}}
\label{fig5}
\end{figure}

\section{Results}
\label{sec:result}

\subsection{Phase diagrams}
\label{sec:diagrams}

\begin{figure}[tbh]
\begin{center}
\includegraphics[scale=0.5]{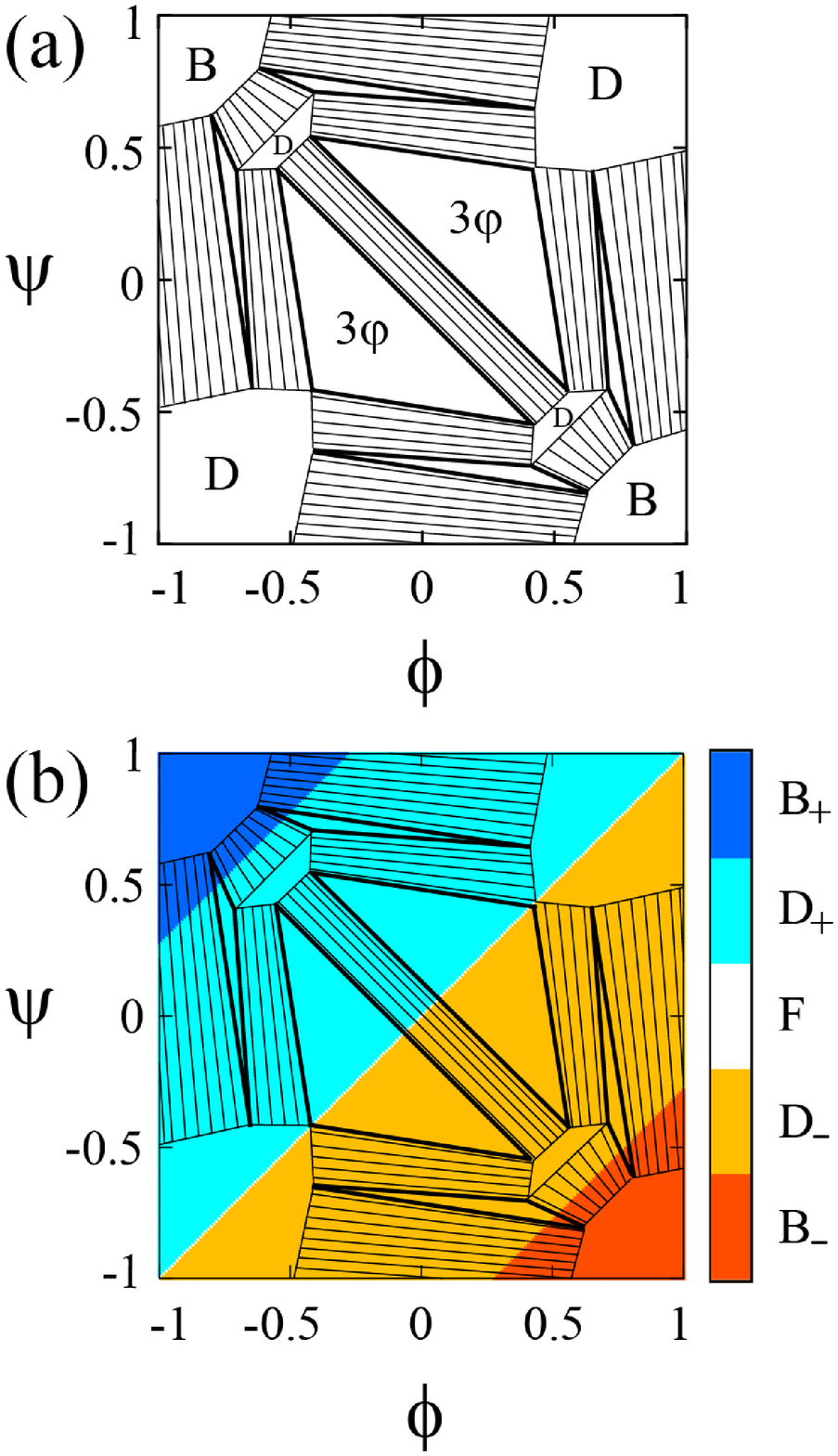}
\end{center}
\caption{\textsf{
(a) Phase diagram in the ($\phi$, $\psi$) plane as in Fig.~\ref{fig4}
but with $t=-0.2$, $\xi=0.25$ and $\beta=1$.
(b) The phase diagram is plotted as in (a) but with a superimposed colored plot for the
curvature, $c$.
 }}
\label{fig6}
\end{figure}

\begin{figure}[tbh]
\begin{center}
\includegraphics[scale=0.5]{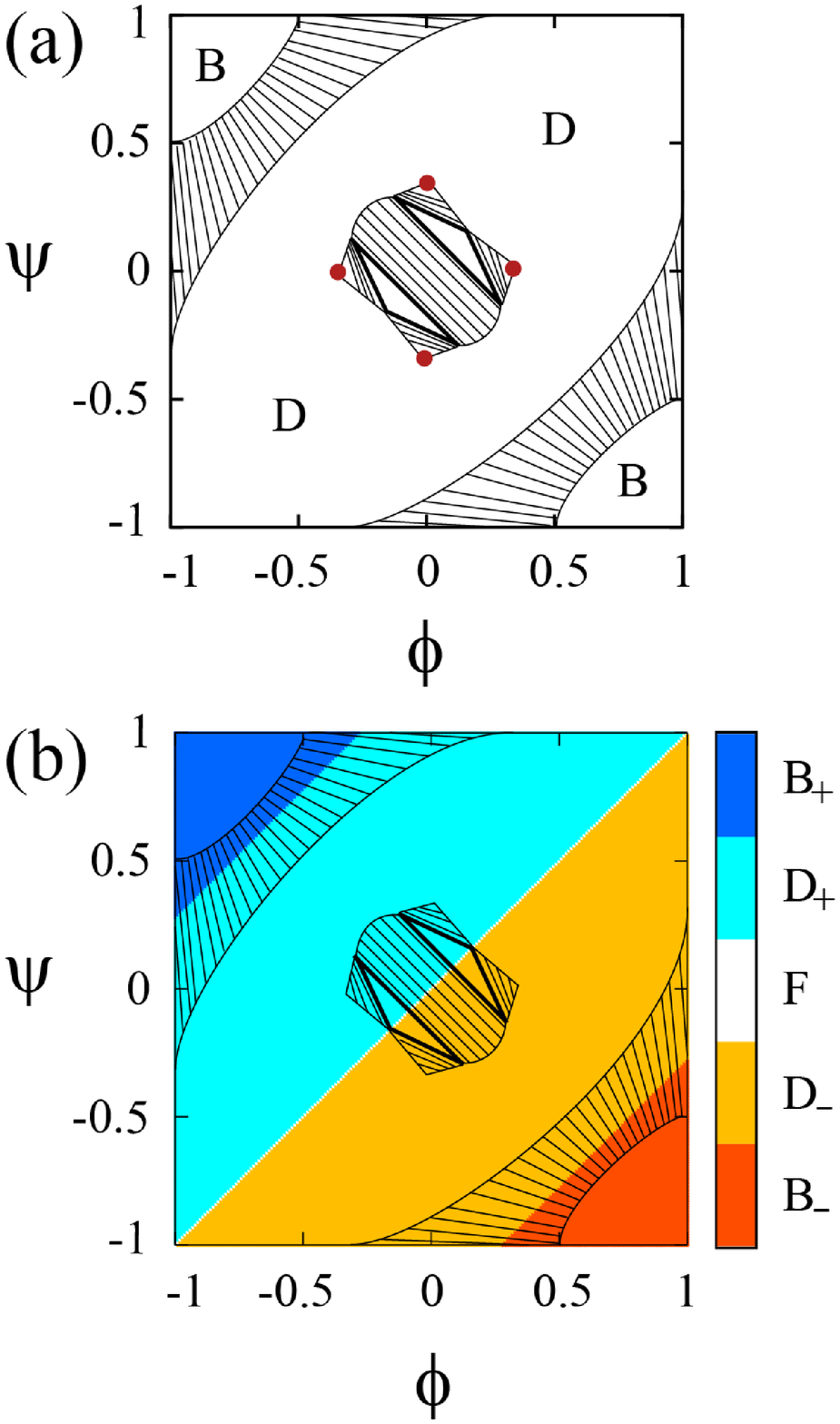}
\end{center}
\caption{\textsf{
(a) Phase diagram in the ($\phi$, $\psi$) plane as in Fig.~\ref{fig4}
but with $t=-0.02$, $\xi=0.25$ and $\beta=1$.
The red circles correspond to the critical points.
(b) The phase diagram is plotted as in (a) but with a superimposed colored plot for the
curvature, $c$.
}}
\label{fig7}
\end{figure}

Four representative types of phase diagrams are shown in Figs.~\ref{fig4}--\ref{fig7}.
In the first two figures, we show the calculated phase diagram for $\beta=1$ (Fig.~\ref{fig4})
and $\beta=3$ (Fig.~\ref{fig5}), while the other parameters are fixed to
$\xi=0.25$ and $t=-0.45$.
The latter two phase diagrams are for $t=-0.2$ (Fig.~\ref{fig6}) and
$t=-0.02$ (Fig.~\ref{fig7}), while keeping $\beta=1$ and $\xi=0.25$.
For presentation purposes, in (b) of each figure, we have superimposed the color plot for
the curvature $c$ on the phase diagram presented in (a).
As denoted above, the various stable phases are the fully-budded phase (B) with
curvature $\vert c\vert=2$, the dimpled or partial budded phase (D) with  curvature
$0<\vert c \vert< 2$, and the flat membrane (F) with  zero curvature $c=0$.
The subscripts $\pm$ denote whether the bud is curved positively or negatively with
respect to the positive normal direction of the planar membrane.

Figure~\ref{fig4} includes five homogeneous phases: B$_\pm$, D$_\pm$ and F.
The B$_+$ and B$_-$ occupy the regions around the $\phi=-\psi=1$ and $\phi=-\psi=-1$ corners, respectively.
On the other hand, the D$_\pm$ occupy the two remaining corners: $\phi=\psi=\pm 1$. The difference between D$_+$ and D$_-$ is only associated with their curvature ($c\lessgtr 0$), which changes continuously.
This is further clarified in part (b) of the figure.
For all $t$-values, the flat F phase (not seen on the figure) strictly  exists only on the $\phi=\psi$ diagonal, but is important for the multi-phase coexistence regions (as discussed below). As one approaches this line from above or below, the D$_\pm$ phases change-over smoothly into the F phase (with $c=0$) on the diagonal line.

Moreover, five two-phase coexistence regions are shown in Fig.~\ref{fig4} together with their calculated tie-lines: two
B$_+$/D$_+$ and two B$_-$/D$_-$ along the boundaries of the phase diagram, and one B$_+$/B$_-$ along the major diagonal, $\phi=-\psi$.
In addition, two three-phase coexistence regions:  B$_+$/B$_-$/F can be seen. These are the two triangular regions lying above and below the $\phi=-\psi$ diagonal. Note that the F corner of the three-phase region lies close to the D$_\pm$ corners, but since it lies on the $\phi=\psi$ diagonal, it is identified as the F phase with $c=0$.
The three-phase coexistence region between F and B$_\pm$ phases means that
each point inside the triangular region is composed of three relative area fractions of the three coexisting phases: the flat (F) and  budded (D$_\pm$) phases.
The nearly horizontal (D$_-/$B$_-$) or vertical (D$_+/$B$_+$) tie-lines on the boundaries of phase diagram indicate that the two $\phi$ and $\psi$ monolayers are almost decoupled, because either $\phi$ or $\psi$ do not vary along the tie-line. On the other hand, tie-lines that lie along the major diagonal, $\phi=-\psi$, indicate a strong coupling between the $\phi$ and $\psi$ monolayers in the B$_+/$B$_-$ coexistence region.

In Fig.~\ref{fig5} with $\beta=3$ and the same temperature as in Fig.~\ref{fig4}, we see that the central binary coexistence region (B$_+$/B$_-$) becomes much larger. On the other hand, the four two-phase coexistence regions of Fig.~\ref{fig4} have
shrunken because the extent of the dimpled phase becomes smaller for larger coupling parameter, $\beta$. When $\beta$ gets large values in Eqs.~(\ref{spontaneous1}) and (\ref{spontaneous2}),
the composition of each monolayer induces higher curvature that promotes budding. In this case, the central two-phase region, B$_+$/B$_-$, occupies a large fraction of the phase diagram, and most of its tie-lines are parallel to the diagonal $\phi = -\psi$, suggesting a strong coupling between the two monolayers. On the other hand, the two three-phase coexistence regions, B$_+$/B$_-$/F, become smaller in Fig.~\ref{fig5}. Furthermore, we note that in Fig.~\ref{fig5}(b), the regions of the dimpled phases,
D$_\pm$, represented by the light blue and orange regions are narrower as compared with those of Fig.~\ref{fig4}(b).

It is of interest to explain how one of the monolayers induces a phase
transition in the second monolayer.
As an illustrative example, Let us consider a point in Figs.~\ref{fig4} and
\ref{fig5} with average leaflet composition, $(\phi,\psi)=(0,-0.9)$.
At these compositions, the bilayer separates in Fig.~\ref{fig4} into a D$_-$ phase with compositions
$(-0.75,-0.9)$ and a B$_-$ phase with $(0.75, -0.9)$.
In this weak coupling case, $\beta=1$, the phase separation in one monolayer
does not induce any instability leading towards phase separation in the second monolayer, because the tie-line is nearly
parallel to the horizontal $\phi$-axis.
On the other hand, in Fig.~\ref{fig5} with a larger value of $\beta=3$, the bilayer separates into a B$_-$ phase with $(0.1, -0.9)$ and a B$_+$ phase with $(-0.9, 0.1)$.
As the tie-line in this case lies along the major diagonal,
$\phi=-\psi$, the two monolayers are influencing each other.
Such a situation results from a strong composition-curvature coupling
when the parameter $\beta$ is large enough.
A more general dependence of the phase diagram on the parameter
$\beta$ will be further discussed in the next subsection.

Figure~\ref{fig6} is plotted for a higher temperature $t=-0.2$ than in Figs.~\ref{fig4} and {\ref{fig5}, while we fix $\beta=1$ as in Fig.~\ref{fig4}. The dimpled region expands both toward the corners and the middle of the phase diagram.
Moreover, there are two new one-phase regions of the dimpled phase (D$_\pm$)
resulting in four additional two-phase coexistence regions:
two D$_+$/D$_+$ and two D$_-$/D$_-$.
The system exhibits a first-order phase transition in composition,
while the transition is second-order in curvature.

In Fig.~\ref{fig7}, the temperature is increased to $t=-0.02$, while $\beta$ and $\xi$ stay as in Fig.~\ref{fig4}. The chosen $t$-value is higher than in the previous figures, and approaches the critical temperature, $t_{\rm c}=0.04$~\cite{Wolff}.
The central region of the phase diagram is dominated by the dimpled phase (D$_\pm$), while the budded regions (B$_\pm$) exist only close to the two corners, with two-phase coexistence regions, B$_+$/D$_+$ and B$_-$/D$_-$.
As seen in (b), the regions of the dimpled phases (D$_\pm$) are determined by $\beta$, and do not depend on the temperature $t$ as long as $\xi$ is fixed.
Furthermore, there are four critical points appearing in the central region (marked by red circles), for which the system exhibits a second-order phase transition both in curvature and composition.

\subsection{Global phase behavior}
\label{sec:global}

\begin{figure}[tbh]
\begin{center}
\includegraphics[scale=0.6]{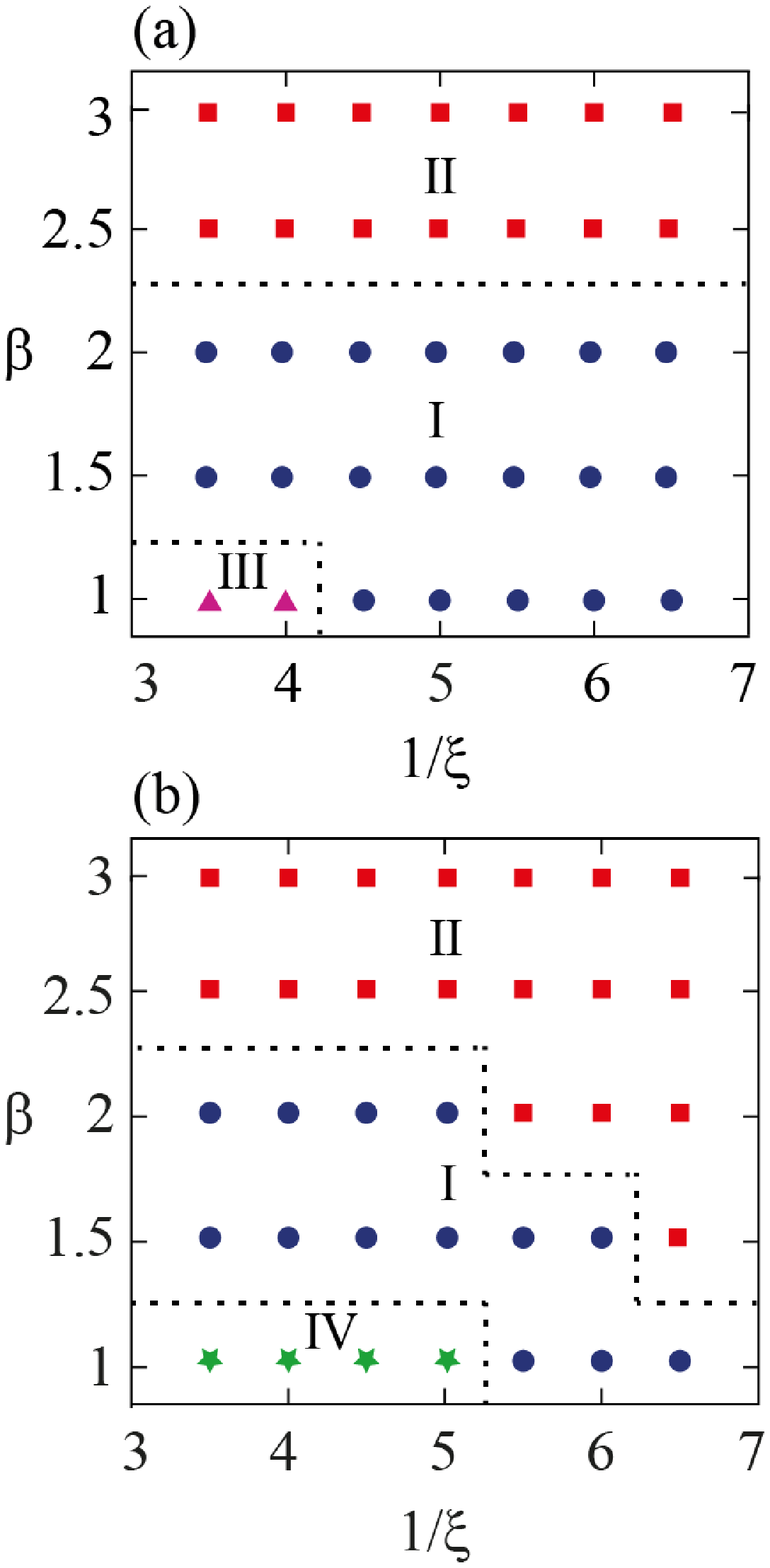}
\end{center}
\caption{\textsf{
Schematic phase-behavior plot in the ($1/\xi, \beta$) plane for temperatures
(a) $t=-0.25$ and (b) $t=-0.02$.
The diagram types, I, II, III and IV, correspond to representative examples
as in Figs.~\ref{fig4}, \ref{fig5}, \ref{fig6}, and \ref{fig7}, respectively.
}}
\label{fig8}
\end{figure}

Next, we investigate how the different phase-diagram types appear and change as we adjust the system parameters in a global way.
In Fig.~\ref{fig8}, we present the global phase behavior in the ($1/\xi, \beta$) plane for $t=-0.25$ in (a)  and $t=-0.02$ in (b).
We show how the four different types of phase diagrams of Figs.~\ref{fig4}, \ref{fig5}, \ref{fig6}, \ref{fig7}, labelled as I, II, III and IV, respectively, evolve as function of $\xi$ and
$\beta$.

Figure~\ref{fig8}(a) summarizes the result for $t=-0.25$ and is valid even for lower $t$-values.
For $\beta>2$ (strong coupling),  the diagram is that of type II in which the dimpled phase (D) region shrinks while the budded phase (B) expands.
For $\beta<2$ (weak coupling), on the other hand, the diagram is mostly of type I for which the wide three-phase coexistence occurs.
For $\beta<1$ and $1/\xi<4$, type III diagram is found, and contains
coexistence regions of the dimpled phases (D$_+$/D$_+$ and D$_-$/D$_-$).

In Fig.~\ref{fig8}(b), for $t=-0.02$ that is closer to the critical temperature, type III is replaced by type IV for $\beta<1$ and $1/\xi<5$, while type II also extends to smaller $\beta$ when $1/\xi$ is large enough.
When the temperature is higher, the phase transition in curvature
from  the flat (F) phase to the dimpled (D)
one, becomes continuous~\cite{Wolff}.

\section{Discussion}
\label{sec:discussion}

In this paper, we have discussed a phenomenological model that accounts for the budding transition of asymmetric two-component lipid domains, where the two domains have in general different average compositions, represented by $\phi$ and $\psi$. Assuming a linear composition dependence of the spontaneous curvature, we have taken into account a coupling between the local curvature and local lipid composition in each of the two leaflets. We then explored the morphological changes between the flat and budded domains by using a thermodynamic argument. Our free-energy model contains three contributions: bending energy, accounting for domain deformation in the normal direction; line tension along the rim of the budded or flat domain;
and a Landau free-energy expansion, which accounts for a phase separation of the two-component lipid domains. We have assumed, in addition, that the domain area remains constant during the budding process.

Our model predicts three different states for the domains: fully budded (B), dimpled (D), and flat (F) states. In particular, in some ranges of parameters, the D state is found to be the most stable one, as observed in the experiment~\cite{RUPK}.
Within mean-field theory, we have calculated various phase diagrams in terms of two compositions for different temperatures $t$, domain size $1/\xi$, and coupling parameter, $\beta$.
The resulting phase behavior is very rich.
The calculated phase diagrams include various one-phase, two-phase (e.g., B$_+$/D$_+$, B$_-$/D$_-$ or B$_+$/B$_-$) and three-phase coexistence regions (e.g., B$_+$/B$_-$/F), depending on the curvature-composition parameter $\beta$ as well as the temperature. Finally, four different types of phase diagram morphologies are found, and we have analyzed the global phase behavior in terms of the coupling parameter $\beta$, and
the domain size $\xi$. The model analysis suggests that the asymmetry in the lipid composition between the two leaflets can lead to complex morphological and thermodynamic behavior of lipid domains.

The most important mechanism that leads to the inter-leaflet correlated phase separation, such as the two-phase coexistence seen in Fig.~\ref{fig5} between two budded phase (B$_+$/B$_-$), is the composition-dependent spontaneous curvature, introduced in Eqs.~(\ref{spontaneous1}) and (\ref{spontaneous2}).
With this mechanism, the coupling between domain curvature and its composition is controlled by the parameter $\beta$. Although we did not include any direct interaction between the domains occurring on the two leaflets, the fact that the two domains are in full registry to each other and have opposite curvatures, $\pm C$, results in a strong correlation between the opposing domains. We consider that such an effect generally exists in bio-membranes, even in the absence of specific proteins that maintain the compositional asymmetry between the two leaflets.

When the coupling parameter $\beta$ is made larger at a fixed temperature
(compare Figs.~\ref{fig4} and \ref{fig5}), the two-phase coexistence region,
B$_+$/B$_-$, dramatically expands due to the coupling between the two leaflets. Such a strongly correlated phase separation can be in general observed for lower temperatures. We also note that the stable budded (B) phase occupies the two asymmetric corners of the phase diagram, i.e., $\phi=-\psi=\pm 1$ . Moreover, the region of the dimpled (D) phase increases as the temperature is raised towards the critical temperature from below (compare Figs.~\ref{fig4}, \ref{fig6} and \ref{fig7}).
Especially, in Fig.~\ref{fig7}, the dimpled one-phase region dominates the central region of the phase diagram and four different critical points are expected to appear. In the intermediate temperature, as in Fig.~\ref{fig6}, the phase diagram contains several types of two-phase coexistence regions.

The present work for asymmetric domains is a generalization of our previous model~\cite{Wolff}, where only the average composition of the domains in the two leaflets is controlled by a single chemical potential.
Here, we have considered a more general situation, where each of the two domain composition is controlled in a separate and independent way. This is done by introducing two independent chemical potentials coupled to the two domain compositions, $\phi$ and $\psi$, as in Eq.~(\ref{grand}).
Hence, each domain has a conserved lipid composition that is independent from the other one. The resulting phase diagrams show a much richer phase behavior, while the previous results~\cite{Wolff} can be recovered by considering the special case of $\phi+\psi={\rm const.}$

Our assumption that the two opposing domains in different leaflets are correlated to each other is in accord with several experimental observations. For studies of planar but asymmetric composition in the two leaflets, Collins \textit{et al.}~\cite{Collins} reported that in some cases, one leaflet can induce a phase separation in the other leaflet, depending on local lipid composition. Whereas in other cases, the two leaflets do not interact. A similar experimental phenomenon was observed for membranes composed of lipids extracted from biological cells~\cite{Kiessling}.
Such situations for asymmetric membranes can be partially described by the different phase behaviors of asymmetric membrane depending on the coupling parameter $\beta$ as shown in Fig.~\ref{fig8}.
Moreover, both in our model and in experiments~\cite{Collins}, domain-induced processes take place in lipid membrane without any proteins.
These results may suggest a cellular mechanism for regulating protein function by modulating the local lipid composition or inter-leaflet interactions.

Although we have mainly discussed the domain-induced budding, our model
can be applied to describe the formation of vesicles in mixed amphiphilic systems~\cite{SH}. It was observed in experiment that mixtures of anionic and cationic surfactants in solution form disk-like bilayers in some range of the relative amphiphilic composition. As these disk-shaped bilayers grow in size, they transform into spherical caps and eventually become spherically closed vesicles. In such cases, the spontaneous curvature of bilayer membranes may be induced due to the compositional asymmetry between the two monolayers.

Finally, our model suggests that the asymmetry in the lipid composition between the two leaflets
leads to a complex behavior of lipid
domains even in the absence of any specific enzymes or proteins, which can induce additional coupling between
the two leaflets~\cite{Gri04}. The importance of such a pure physical mechanism can be verified in experiments on asymmetric model membranes involving only a lipid mixture. They also
can be of relevance to signal transduction~\cite{Simons00},
membrane fusion~\cite{Puri06}, or penetration of viruses into cells~\cite{Chazal03}. We hope that
additional experiments will address these issues in the future.

\bigskip
{\em Acknowledgments.~}
J.W.\ acknowledges support from the Service de Coop\'eration Scientifique et Universitaire de
l'Ambassade de France en Isra\"el, the French ORT association, and ORT school of Strasbourg.
S.K.\ acknowledges support from the Grant-in-Aid for Scientific Research on
Innovative Areas ``\textit{Fluctuation and Structure}" (Grant No.\ 25103010) from the Ministry
of Education, Culture, Sports, Science, and Technology of Japan, and
the Grant-in-Aid for Scientific Research (C) (Grant No.\ 15K05250)
from the Japan Society for the Promotion of Science (JSPS).
D.A. acknowledges the hospitality of the KITPC and ITP, Beijing, China, and partial support from
the Israel Science Foundation (ISF) under Grant No.\ 438/12,  the United States--Israel Binational Science Foundation (BSF) under Grant No.\ 2012/060, the ISF-NSFC joint research program under Grant No.\ 885/15, and a CAS President's International Fellowship
Initiative (PIFI, China).

\appendix*
\section{The Hessian and Stability Analysis}
In order to discuss the stability of the free-energy, we consider the
$2\times 2$ Hessian matrix of $\varepsilon^{\ast}(\phi,\psi)$ given by
\begin{eqnarray}
\mathbf{H} =\left(
\begin{array}{cc}
\partial_{\phi\phi}\varepsilon^{\ast} &  \partial_{\phi\psi}\varepsilon^{\ast}  \\
\partial_{\psi\phi}\varepsilon^{\ast}  &  \partial_{\psi\psi}\varepsilon^{\ast}  \\
\end{array}
\right),
\label{eq:bb}
\end{eqnarray}
where $\partial_{ij}\varepsilon^{\ast}$ are the second-order partial derivatives of $\varepsilon^{\ast}$. We recall that by minimizing $\varepsilon(\phi,\psi,c)$ with respect to $c$, we obtained $\varepsilon^*$
as in Eq.~(\ref{epsast}). This leads to $\partial_c \varepsilon=0$,
$\partial_{\phi} \varepsilon^{\ast}=\partial_{\phi} \varepsilon$,
and $\partial_{\phi} c=-\partial_{c\phi}\varepsilon/\partial_{cc} \varepsilon$, as well as similar expressions
for the $\psi$ derivatives. Using these relations, one can show that the components of $\mathbf{H}$ are given by
\begin{align}
&\partial_{\phi\phi}\varepsilon^{\ast}=
\partial_{\phi\phi}\varepsilon-\frac{(\partial_{c\phi}\varepsilon)^{2}}{\partial_{cc}\varepsilon}, \nonumber \\
&\partial_{\psi\psi}\varepsilon^{\ast}=
\partial_{\psi\psi}\varepsilon-\frac{(\partial_{c\psi}\varepsilon)^2}{\partial_{cc}\varepsilon}, \nonumber \\
&\partial_{\phi\psi}\varepsilon^{\ast}=
\partial_{\psi\phi}\varepsilon^{\ast}=
\partial_{\phi\psi }\varepsilon-\frac{(\partial_{c\phi}\varepsilon)(\partial_{c\psi}\varepsilon)}
{\partial_{cc}\varepsilon}.
\end{align}
At the spinodal condition, the Hessian defined as the determinant of the matrix, $H=\det \mathbf{H}$, vanishes. This condition can be written as
\begin{align}
&\left[2\beta^2+\frac{1}{\xi^2}(t+3\phi^2)-\frac{4\beta^2}{\partial_{cc}\varepsilon}\right] \nonumber \\
& \times \left[2\beta^2+\frac{1}{\xi^2}(t+3\psi^2)-\frac{4\beta^2}{\partial_{cc}\varepsilon} \right]
-\frac{16\beta^4}{(\partial_{cc}\varepsilon)^2}=0,
\label{eq11}
\end{align}
where
\begin{equation}
\partial_{cc}\varepsilon=4-\frac{1}{4\xi (1-c^2/4)^{3/2}}.
\end{equation}

The critical point can be derived by considering another $2\times2$ matrix
\begin{eqnarray}
\mathbf{H}' &=&\left(
\begin{array}{cc}
\partial_{\phi\phi}\varepsilon^{\ast} &  \partial_{\phi\psi}\varepsilon^{\ast}  \\
\partial_{\phi} H  &  \partial_{\psi} H  \\
\end{array}
\right),
\end{eqnarray}
and its determinant $H'=\det \mathbf{H}'$. One can explicitly show that
\begin{align}
\partial_{\phi} H &=
(\partial_{\phi\phi\phi}\varepsilon^{\ast})(\partial_{\psi\psi}\varepsilon^{\ast})-
(\partial_{\phi\psi}\varepsilon^{\ast})(\partial_{\phi\phi\psi}\varepsilon^{\ast}) \nonumber \\
&+(\partial_{\phi\phi}\varepsilon^{\ast})(\partial_{\phi\psi\psi}\varepsilon^{\ast})-
(\partial_{\phi\phi\psi}\varepsilon^{\ast})(\partial_{\phi\psi}\varepsilon^{\ast}),
\end{align}
\begin{align}
\partial_{\psi} H &=
(\partial_{\phi\phi\psi}\varepsilon^{\ast})(\partial_{\psi\psi}\varepsilon^{\ast})-
(\partial_{\phi\psi}\varepsilon^{\ast})(\partial_{\phi\psi\psi}\varepsilon^{\ast}) \nonumber \\
&+(\partial_{\phi\phi}\varepsilon^{\ast})(\partial_{\psi\psi\psi}\varepsilon^{\ast})-
(\partial_{\phi\psi\psi}\varepsilon^{\ast})(\partial_{\phi\psi}\varepsilon^{\ast}).
\end{align}
Then, the conditions for the critical point are given by~\cite{Gibbs}
\begin{equation}
H=0~~{\rm and} ~~H'=0.
\label{cri}
\end{equation}


\end{document}